# Redefine the correlation coefficient by experiment methods

Xiatong Cai[1], Guangpeng Pei[1,2], Yuen Zhu[1], Donggang Guo[1]*, Hua Li[1]*

[1] School of Environment Science and Resources, Shanxi University, Taiyuan, Shanxi 030006, China.

[2] Institute of Resources and Environment Engineering, Shanxi University, Taiyuan, Shanxi 030006, China.

**Abstract**: With the establishment of global biological monitor network and development of remote sensing technology, data won't be a limitation, but the variance brought by spatial heterogeneous and fractal will influence correlation coefficient significantly with the enlarged sample scale. Those impede us to find more intrinsic principle in ecology. Ecology is based on experiment, and the experiment methods won't change with spatial. In that condition, if we construct a system to evaluate the experimental difference, that may benefit the study in spatial ecology. In that condition, we give a synthesis discussion from the concept of data, experiment to analysis methods, and revise as well as develop correlation and regression method to make it suit the demand of spatial analysis. An experimental correlation system is established in this report.

---

* Corresponding Author:
Hua Li, School of Environmental Science and Resources, Shanxi University, Taiyuan 030006, China; Tel: +86-351-7018782; Fax: +86-351-7010600; Email: lihua@sxu.edu.cn;
Donggang Guo, School of Environmental Science and Resources, Shanxi University, Taiyuan 030006, China; Tel: +86-351-7018782; Fax: +86-351-7010600; Email: guodongg@sxu.edu.cn;

**Introduction**

The rapid global climate change is altering some environmental processes (IPCC 2015), such as causing the boundary shift of some species (Hargreaves et al. 2014, Alexander et al. 2015), accelerating soil microbe activities (Crowther et al. 2015), and changing the biodiversity at global scale (Sala et al. 2000, Urban et al. 2016). To grasp the global change influence, macroecology gives many potential solutions to study the global change (Kerr et al. 2007), and by the help of the developing technology such as remote sensing (Turner et al. 2003) and more economic support (Mervis and Kaiser 2003), researches could make analysis on a large scale.

However, combining with large scale data, several questions are also brought forward, especially the spatial heterogeneous, combining with different ecological pattern, which leads to the result change with the sample scale change (Levin 1992, Dutilleul 1993, Johnson et al. 2010). For example, although the microbial organic nutrient acquisition in soil and sediment show a particular patter, the pattern in those two types of area are different (Sinsabaugh et al. 2009), and even if the atomic ratio of C: N: P on soil and soil microbial are similar at global scale, significant variances are still observed by researchers (Cleveland and Liptzin 2007). Similarly, results are also observed at some animal studies (Kerr and Packer 1997, Johnson et al. 2010). A problem found in macroecology is that the correlation and the regression results usually lower than we thought (Schlesinger et al. 1996, Storch et al. 2005, Fierer et al. 2009). This pushes us to revise original methods and establish more suitable methods to cope with this phenomenon.

The data in spatial will vary a lot with the land types and scale sampled, but the experiments' methods won't. That inspire us to think whether we could develop a new evaluation method to measure the difference between two experiments methods. Here, we are aiming to develop new concepts from data, experiments to mathematical methods and give answer to two questions: (1) why the correlation and regression method is not suitable in the future ecological spatial analysis? (2) How could we establish a system that could be used to measure the relation between two relatives? For a convenient illustration, we will arrange discussion from soil ecology aspects.

**Observation and the information.**

It is desirable to call the experiment the observation and the experiment's result the information. I want to compare the observation to look a set of information at a certain angle. From the mathematical point, it is more likely to compare to map a hypergeometric to a number, and different information, such as the soil organic carbon and the nitrogen, is to observe the hypergeometric from different angle. The reason we say that is because we can never get a single information by an observation. For example, the organic carbon is a synthesis concept, because it contains decaying vegetation, bacterial, and metabolic activities of living organisms or chemicals, and the soil microbial biomass means the synthesis of all the microbes in the soil (Wardle et al. 2004). So that the intrinsic meaning of correlation is to evaluate how close the two synthesis concepts and how overlap or different they are.

**A discussion on the mathematical methods**

*Correlation and regression*

It is interesting to find the relationship between two different 'organs' and sometimes amazing that the results even seems weird at first sight but finally we find reason. The correlation method, which analyzes the relationship between two random variables or bivariate and values the relationship from vague description such as 'high', 'medium' or 'low' to quantity number from -1 to 1. Since Galton published his great work-*Natural Inheritance* (Galton 1889)- the statistic went into a huge development (Pearson 1896, Spearman 1904) and more and more statistical methods come out based on their study.

For discussion, we firstly suppose the influence from spatial heterogeneous could be ignore in an enough fine area. That is reasonable because although different areas share different ecological patterns (Watt 1947), the autocorrelation could make sure there is a fine scale that different area share a similar pattern even though it is a trouble to analyze some questions(Legendre 1993, Marie-JoseFortin and Dale 2005). With the scale enlarged, more geological land types will be collected in, and may influence the result. To give a clear illustration, a set of data are used here. The data we used comes from the thesis of Cleveland (Cleveland and Liptzin 2007). He summarized the soil nutrients and the microbial biomass of most researches and classified it into three categories. We choose the latitude, soil carbon and the soil nitrogen in his article, eliminate the missing values and divide it into two categories by the latitude (0º~30º, and 30º~70º). The Spearman correlation coefficient has been applied to each category and both data by using SPSS 19.0. Figure.1 shows the result that different category has different correlation coefficient and the low latitude's coefficient is smaller than the high latitude. That could be explain by the different vegetation type, because at low latitude, the experiment manipulated mostly in the forest, but the high latitude was on the grassland. (We choose 30º as the boundary of two categories, because the data at latitude 27º shows a different pattern, and we category it to the lower latitude set, considering of the experiment at lower latitude mostly applied at the forest.) The total data's coefficient falls between the result at the low latitude's coefficient and the high latitude's result, which means that the coefficient could be influenced by different process, when we expand the collected scale. The similar results are observed on the Karr's paper, too (Kerr and Packer 1997). We would like to call this phenomenon *processes neutralization*. So, we could say that, even the correlation coefficient gives a useful feedback to the relationship between different information, but it will be affect by the process enrichment. Traditionally, more data mean the result contains more information and the result will be more accurate, but as for the spatial ecology and to our special question, that may loss its meaning.

It is irresponsible if we just left it there and superficially think that this effect is a natural characteristic and if we know different landscape process, we still could predict the natural process by more data acquired and calculating the result in different land type by different models. Giving the scale transition mechanism a reasonable explanation, not only could it make us have a more accurate model to predict the global change, but, the most important thing, to know what will the system change under the pure scale change. Combined with Gaia theory (Lovelock and Margulis 1974), only if we know the mechanism that ecological process changed with the pure scale, we can know the organisms' organization from cell to biosphere. Therefore, we couldn't ignore the importance to study the scale effect.

A similar situation happens to the regression too. Galton and Pearson (Galton 1889, Pearson et al. 1903) found some exclusion to the Mendelian' gene theory on characteristics of human eye's color and the dogs' hair color. The Galton brought forward the original thinking about the regression, then Pearson developed it and gave it a mathematical form. They meant to answer the heredity questions which don't accord with the Mendelians. In Pearson's article- *The law of ancestral Heredity*- he gave a mathematical formation of the regression which on the base of correlation analysis. The relationship between two relatives, *P* and *Q*, was been developed into *Q* with multiple variables that *P₁, P₂, P₃…Pₙ,* and the formula like this:

$$p_q = J_1 \frac{\sigma_q}{\sigma_{p_1}} h_{p_1} + J_2 \frac{\sigma_q}{\sigma_{p_2}} h_{p_2} + J_3 \frac{\sigma_q}{\sigma_{p_3}} h_{p_3} \cdots J_n \frac{\sigma_q}{\sigma_{p_n}} h_{p_n} \qquad (1)$$

where $p_q$ is the most probable deviation of *Q*, the *h* is the observed deviations of *P*, σ is the variance, and *J* are known expressions involving only the correlation coefficient. Therefore, regression will be affected by the multiple processes, too. We would like to give an illustration by another way, that to value whether the least-square method could be used to give a solution, because the least-square is a standard approach for the modern regression and fitting method. To answer this question, we need to induce a new concept in this paper named *information overlap*, which comes from the synthesis information observation. Due to the observation get a synthesis information, we can't say the results we get comes from which single concept, the information overlap will cause a multiple correspond, which is like a photomultiplier that $v_1$ corresponds to several results in $v_2$, and those points of $v_2$ inversely corresponds to more $v_1$ points. This phenomenon makes our conclusion unfair. That situation needs us to think whether we still can use residual as the data to adjust the least-square method. We think the residual in the regression shouldn't be defined as the distance that a point to variable's expectation, but should be defined in consider of the distribution of a variable to the other. To specific illustrate this relationship, we use a notation that:

$$\{(.)| ... : B_{[.]}: A_i\} \tag{2}$$

That is a corresponding choice and manipulation notation. We need to announce that this concept should distinguish with the mapping, we use the corresponding relationship, and it comes from the observation and information because we don't know the exact mapping rule, if we use the mapping concept, it will add more uncertainty and that is contrary to our original intention. The notation |.} means a data set. $B_{[.]}: A_i$ means the B points which corresponds to the $i^{th}$ points at variable A. If [.] takes [n] means all the points corresponds to the $i^{th}$ point in A, and if [.] takes [j] means the $j^{th}$ value at $|B_{[n]}: A_i\}$. The '…' means this picking manipulation could be proceed going by adding more pick rule C, D, E… , and the calculation order is from right to the left. The {(.)| means the calculation manipulation, for example, {mean| means calculate the mean value in the

points set |.}. According to this notation, the point $A_i$'s residual($e_i$) could be defined as:

$$e_i = \sum_{j,k}^{n} \{\{Expected|A_k:B_j:A_i\} - \{Expected|A_n:B_n:A_i\}\}^2 \quad (3)$$

The reason we defined the residual in this format is because the information asymmetry, that a value in $v_2$ doesn't exactly correspond to a value in $v_1$, so that the residue by the format like the traditional one may enlarge the uncertainty, however, if we define the residual value in one identical variable set, this problem won't happen. In that condition, the fitting becomes let the sum of $e_i$ least. The spatial autocorrelation can make sure our result could find.

### *The surface fitting method in ecology*

Till now, we haven't given the method suitable to analyse the scale pattern effect or refined the weakness of correlation when it was used to analyse the scale pattern effect. But in some degree, we have found the reason, why the correlation and regression can't be used to solve the problem exist in the spatial. The reason should belong to: (1) Process neutralization effect, which means when pattern types increased with the increased scale the correlation coefficient will be affected by the process. (2) The information overlap effect, which means that the corresponding situation makes the information asymmetry, or the corresponding rule unfair.

We could find that all problems are caused by the information losing, or we call this phenomenon the information deficiency. We think this problem can only be solve by the information expansion, which means diminish the overlap and find enough observation information to get individual distinguish. The phrase, find enough information, doesn't imply to increase the sample data but to make an information coalition to avoid the data undistinguished. One way, we think, is to use the spatial coordinate, because the spatial coordinate is a common factor, and it can reduce the overlap by giving each information a unique description so that every information by an observation could be distinguished. This could also bring more interesting conclusions. In this section, we will discuss how to use the spatial information to build our system. Even though nowadays the existing spatial fitting method can't satisfy our demand. We do an experiment on a natural forest reservation and collected its information on plants distribution, soil carbon, nitrogen, phosphorous as well as the microbial biomass and use the surface polynomial fitting method hoping to test whether our method could have a considerable result, but the fitting results are bad. But by accident, we find an interesting phenomenon that when fitting the soil organic data, although we have poor fitting result, different times of the polynomial fitting equation may have a similar distribution pattern with a kind of vegetation in forest. That may give us an speculation that a species of vegetation distributed in the forest could be represented by a group of soil nutrient functions. However, this phenomenon hasn't been find at all scale, but only when the influence of geological factor become significant, and it is hard to believe the present pattern could be decomposed by several functions, because the interspecific relationship such as the competition and symbiosis make function hard to be represent by some function's combination like the quantum mechanism's formation, because the interspecific relationship will make the system become too complex to describe. In that condition, we don't use this data but just post it as a question here. On top of that, Fourier decomposition and the wavelet decomposition may make the format of the final equation too complex and there still need the fitting method follows with some characteristics and we will discuss it latter, so that those two methods may can't give us a useful answer. We don't know how to get a satisfied fitting result, but we believe we can get and hope there is a function which satisfied our demanding and could be find in the future. (The subsequent discussing is based on there is a function could satisfy our demand.) Due to our narrow horizon, we can't give a proof here, so we hope the further researchers could proof or reject our announcements to give a method to value the different between two observations more systematically.

Before illustrating our thinking, we should make some assumptions:

*Assumption 1*: We could use a continues function (spatial function) to fit the value of variable in the space.

*Assumption 2*: The function is positive and the variable could be separated from the function.

*Assumption 3*: For a function of an observation, its integral is limited and the inner product of two functions exist.

To expand the information, we need to use the spatial fitting method to fit the data set. We suppose we could use two coordinates to represent the global situation, that longitude($x$) and latitude($y$) and we want to use the microbe($m$), soil carbon($c$), soil nitrogen($n$), soil phosphorous($p$) to analysis the coupling effect of the biogeochemical circulation effect. If we can get four functions by the

spatial fitting:

$$f_n(x, y, w) = 0 \quad (4)$$

where *n* represent $n^{th}$ function in the function group and *w* is one soil nutrient. So for a general purpose, we could use the *f₁* (function of *c*) over *f₂*(function of *n*), and *f₁* over *f₃* (function of *p*) to get two function which contain the *f(x,y,c,n)=0* and the *f(x,y,c,p)=0*. Those two function could eliminate the variable *x* to get a function *f(y,c,n,p)=0* . We could also use the function *f₂, f₃* and *f₂, f₄* (function of m) to get a function *f(y,n,p,m)=0,* and if those two *x* eliminated function could eliminate the coordinate *y*, then we get a function about *f(c,n,p,m)=0*. Which is a useful function to learn the coupling pattern about the microbe and the environmental factors. This function could be used to analyze the niche theory such as to talk about the axes characteristic harbor in the Hutchinson's Concluding remark (Hutchinson 1957, Holt 2009), which announced to use the axes to represent the limitations in niche and descript the species on a hypergeometric space.

*A new correlation coefficient*

At former section, we give the concept about the 'angles' between two observations. In that condition, we need to think how to define a new correlation coefficient. We couldn't deny that it is convenient to use [-1,1] coefficient format, so we want to keep this format and combine with the 'angle' concept. We think if *r*=1 means two observations have same observation direction, and *r*= -1 means we observe the buck information at a verse direction. However, that definition is real abstract, if we just use two mapping to define the correlation coefficient and we don't know what is the rule of the mapping as well as the characteristic of the set information. Therefore, it is reasonable to seek help to the spatial function. We do that because the information in spatial equation is enough to describe our object and the coalition information include the geological factor could reflect the scale information.

We need to construct an operator which act on two spatial function and get a number. If the operator could map the two functions into a special number, then the operator must could combine those two functions with a quantity number. Compare the spatial function with the data that doesn't involve the spatial coordinate, the distribution of nongeological variable still be normal distribution, because the data of *x* and *y* are random collected. So according to the original assumption of the Pearson coefficient, the maximum likelihood appears at the same point that (Pearson 1896). However, in prior definition, there is information deficiency phenomenon. Hence, we consider the original meaning of σ. The meaning of σ is aimed to use a constant to represent an animal's trail. In our spatial function, the information of a trail is represented by the function, so we replace σ with the spatial function, and according to the fundamental inequation, the correlation coefficient should be defined as:

$$r_{12} = \frac{<f_1|f_2>}{\|f_1\|\|f_2\|} \tag{5}$$

where the $<.|.>$ means the inner product of two function, $\|.\|$ means the norm of a function, and *r₁₂* means the difference between two observation. The difference between this correlation coefficient with the former is that: (1) the information we expand or the information coalition could represent the data characteristic in the spatial analysis. (2) The coefficient represents the difference between different observation but it could be seen as a correlation coefficient by our former discussion.

**Prospect and limitation**

The method, to use experiment defining the correlation coefficient, give us a new idea to study the spatial scale effect in ecology. However, the format of spatial function is still dimness. That restrains the theory's development. We hope the experiment differences could be represented by a format like the rotation, and positive correlation is clockwise while negative correlation is anticlockwise, by a visual description. That means if we know two experiment's (A and B) correlation coefficient and know a new experiment's (C) correlation coefficient with former A or B. Then the correlation between A and C or B and C could be calculated. But that mostly seems impossible within our study, and the process is to complex to build up except we do enough experiments and enrich the relation between each experiment and summarize it into a sheet to acquire more information about the experiment. In this paper, we give a developed discussion on (1) why we should establish a correlation method based on the experiment methods but not the experiment's results. (2) why the former definition of correlation has a limitation when it is applied to study the scale effect. (3) how could we establish a system based on experiments. We don't think our method conflicts with the ecological statistic method, because although we expand the information to make distinguish, we still hope to know the general rule in nature. We aim to make sure the processes difference in scale won't influence our final result. We hope this idea will be advanced by more and more researcher's researches.

**Acknowledgements**

This work was supported by National Natural Science Foundation of China (31400358) and Science and Technology Special Project of Shanxi Province, China(20121101009). We are thankful to Professor Hong Geng, at Institute of Environment Science, Shanxi University, for his help.

**Figure legends**

**Figure. 1:**

The figure plots the scatter diagrams for soil carbon and nitrogen data, which is arrange by Cleveland and Liptzin (2007), in different latitude category. (a) Plot the overall scale (the latitude ranging from 0° to 70°), and the correlation coefficient(r) is 0.895, $p < 0.01$. The two ellipses are plotted to represent our two categories (latitude from 0° to 30° and 30° to 70°). (b) shows the scatter diagram at the range of lower latitude, from 0° to 30° (r=0.840, $p<0.01$). We could find two obvious patterns at figure 1(b) that may be caused by different process in the soil. Figure (c) is the higher latitude coefficient and the scatter diagram also shows several patterns even it is not more obvious than (a). (r=0.955, $p<0.01$). Two straight lines in (b) and (c) mean to show the different patterns.

**Figure 1**

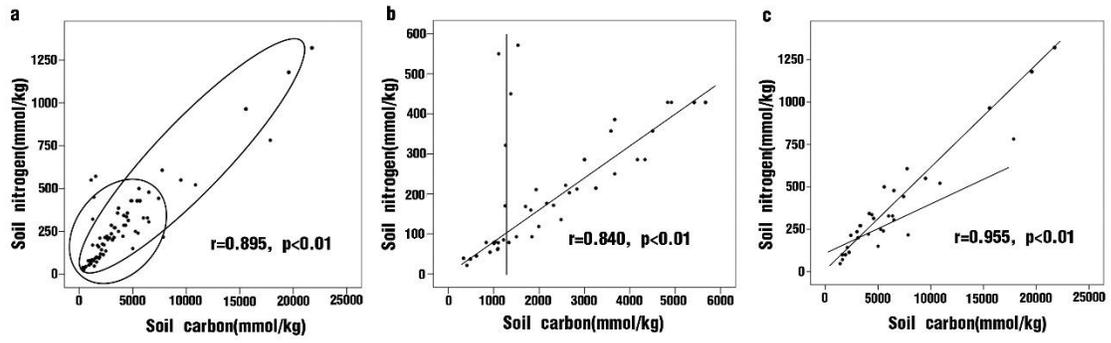